\documentstyle[12pt,aaspp4,epsfig]{article}
\newcommand{\saxj}	{\mbox{\rm\,SAX\,J1808.4-3658}}
\newcommand{\rxte}	{{\em RXTE}}
\newcommand{\bsax}	{{\em BeppoSAX}}

\newcommand{\ginga}	{{\it Ginga}}
\newcommand{\xray}	{\mbox{X-ray}}
\newcommand{\xrays}	{\mbox{X-rays}}
\newcommand{\aprx}	{\mbox{$\sim$}}

\newcommand{\Msun}	{\mbox{$\rm\,M_{\mathord\odot}$}}
\newcommand{\degree}	{\hbox{$^\circ$}}
\newcommand{\lumin}	{\mbox{$\rm\,ergs\,s^{-1}$}}
\newcommand{\gtsim}{\lower.5ex\hbox{$\; \buildrel > \over \sim \;$}}
\newcommand{\etal}              {{\it et~al.\ }}

\newcommand{\onee}	{\mbox{\rm\,1E~1740.7--2942}}
\newcommand{\Grs}	{\mbox{\rm\,GRS~1758-258}}
\newcommand{\ltsim}{\lower.5ex\hbox{$\; \buildrel < \over \sim \;$}}

\begin{document}

\title{The X-ray Spectrum of SAX~J1808.4-3658}
\author{W. A. Heindl}
\affil{Center for Astrophysics and Space Sciences, Code 0424, University of
California, San Diego, La Jolla, CA 92093}
\author{D. M. Smith}
\affil{Space Sciences Laboratory, University of California, Berkeley,
Berkeley, CA 94720}

\authoremail{wheindl@ucsd.edu}

\begin{abstract}

We report on the \xray\ spectrum of the 401~Hz \xray\ pulsar and
type~I burst source \saxj\, during its 1998 April/May hard
outburst. The observations were made with \rxte\ over a period of
three weeks.  The spectrum is well-described by a power law with
photon index $\rm 1.86 \pm 0.01$ that is exponentially cut off at high
energies.  Excess soft emission above the power law is present as well
as a weak Fe-K line.  This is the first truly simultaneous broad-band
(2.5--250~keV) spectrum of a type~I burst source in the hard
state. The spectrum is consistent with other hard state burster
spectra which cover either only the soft (1--20\,keV) or hard (\gtsim
20\,keV) bands, or cover both, but not simultaneously.  The cut-off
power law spectrum resembles that of black hole candidates (BHCs) in
their low states, observed with \rxte.  We compare the \saxj\ spectrum
to three BHCs and find that the power law is somewhat softer. This
suggests that the photon index may provide a way to distinguish
between low state emission from Galactic black holes and type~I
bursters.

\end{abstract}


\section{Introduction}

\saxj\ is the first object thought to display both type~I \xray\
bursts and coherent \xray\ pulsations.  Its low implied magnetic field
(B \ltsim$\rm 2 \times 10^8$\,G, Wijnands \& van der Klis 1998b) and
high spin frequency may make \saxj\ a missing link in the evolution of
the millisecond radio pulsars. It was discovered in observations of
the Galactic center region made during 1996 September 12--17 with the
Wide Field Camera (WFC) on \bsax\ (\cite{intZ98}).  During six days of
observations, the flux level was \aprx\ 50--100\,mCrab (2--10\,keV).
Earlier and later observations which did not detect the source limit
the outburst duration to between 6 and 40 days. This duration was
confirmed with data from the All Sky Monitor (ASM) on the {\em Rossi
X-ray Timing Explorer} (\rxte), which detected \saxj\ for about 20
days beginning 1996 September 8 (\cite{intZ98}).  Two type~I \xray\
bursts were also detected during the \bsax\ observations, making the
identification of the source as a neutron star in a low mass \xray\
binary highly probable.  Assuming that these bursts reached the
Eddington luminosity for a 1.4\Msun neutron star implies a distance of \aprx
4\,kpc.

\begin{figure}
\caption{The light curve of the 1998 April outburst of \saxj\ in three
energy bands from the ASM, PCA, and HEXTE on \rxte. The ASM data are
the publicly available daily averages provided by the ASM/RXTE team
(http://heasarc.gsfc.nasa.gov/docs/xte/asm\_products.html).\label{f_lc}}
\centerline{\epsfig{file=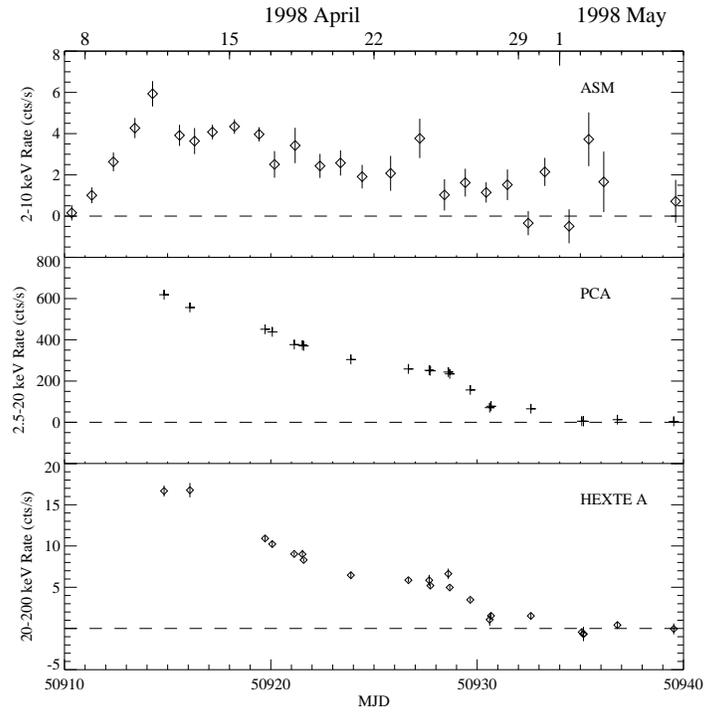,width=5.4in}}
\end{figure}

Following the 1996 outburst, \saxj\ remained undetected until a slew
of the \rxte\ pointed instruments on 9 April 1998 serendipitously
detected a source (designated XTE~J1808-3658) whose location is
consistent with the \bsax\ error region (\cite{Mar98}).  The flux level
at this time was \aprx 50\,mCrab (2--10\,keV), corresponding to a
luminosity of $\rm 1.5 \times 10^{36}$\,ergs/s at a distance of
4\,kpc.  Twenty-one \rxte\ pointed observations over the next 4~weeks
saw the flux increase to 60\,mCrab (2.5-20\,keV) and decrease approximately
exponentially with a time constant of about 10~days (see
Figure~\ref{f_lc}).  After 26~May, the source dimmed rapidly by a
factor of \aprx 5 in 2 days \nocite{Glf98}.

Timing analyses of \rxte/Proportional Counter Array (PCA) data from 11
April 1998 revealed that \saxj\ is an \xray\ pulsar with a frequency
of 401\,Hz, making it the first accretion-powered millisecond \xray\
pulsar (\cite{Wij98a,Wij98b}).  The pulsed amplitude was quite low,
only \aprx 4\% RMS (2--60\,keV). Chakrabarty \& Morgan
(1998a)\nocite{Cha98a}, also using PCA data, detected the binary
orbit. They derived an orbital period of
7249.119(1)\,s, a projected semimajor axis of $\rm a_xsin{\em i} =
62.809(1)$\,lt-ms, and an \xray\ mass function of $\rm 3.85 \times
10^{-5}$\Msun (\cite{Cha98b}).  They also placed upper limits on the
pulse frequency derivative and the eccentricity ($< 5\times 10^{-4}$)
of the orbit.  The very small mass function implies that the companion
mass is $\rm <0.18$\,\Msun\ for a neutron star less massive than 2\,\Msun\
(\cite{Cha98b}).

During the recent \xray\ outburst, optical imaging of the \saxj\ field
revealed an object with magnitude $\rm V = 16.6$ that was not present
in the Digitized Sky Survey (DSS) to a limiting magnitude of $\rm V >
19 $ (\cite{Roc98}).  This object was confirmed as the likely optical
counterpart of \saxj\ when multiple V-band exposures covering the
2\,hr binary orbit showed ``roughly sinusoidal'' variability of
0.12\,mag (\cite{Gil98}).

Early work with the \rxte\ data indicated that the \saxj\ spectrum was
Crab-like and continued unbroken to energies greater than 100\,keV
(\cite{Hei98,Gil98}.  In this letter, we perform detailed spectral
studies and show that the spectrum is somewhat harder than the Crab at
low energies (\ltsim 30\,keV) and is exponentially cut off at higher
energies.

\section{Observations and Analysis}

The recent outburst was the subject of a monitoring campaign with the
PCA and the High Energy \xray\ Timing Experiment (HEXTE) on \rxte.
The PCA (\cite{jah96}) is a set of 5 Xenon proportional counters
(2--60\,keV) with a total area of \aprx 7000\,$\rm cm^2$. The HEXTE
consists of two clusters of 4~NaI(Tl)/CsI(Na) phoswich scintillation
counters (15--250~keV) totaling \aprx1600\,$\rm cm^2 $ area
(\cite{gru96,rot98}).  The two clusters alternate rocking between
source and background fields to measure the background.  The PCA and
HEXTE share a common 1\degree\ full width half maximum field of view.
In this {\em Letter} we discuss the \xray\ spectra obtained during 13
observations made between 1998 April 11 and 25.  Figure~\ref{f_lc}
shows the flux history of \saxj\ in three energy bands.  The outburst
peaked between April 10 and 12 and declined over the next 20~days.

For each observation, we accumulated PCA and HEXTE spectra without
regard to the binary orbit and the \xray\ pulse phase.  These spectra
are quite hard, being superficially similar to the Crab Nebula and
pulsar (a power law with photon index of \aprx 2) and extending to
over 100\,keV (\cite{Hei98,Glf98}).  The PCA background was estimated
using PCABACKEST version 1.5, with the background model based on
blank-sky pointings. We then used XSPEC to fit various models to the
observed counts spectra. PCABACKEST and XSPEC are standard NASA
software tools.  In all fits the relative normalizations of the PCA
and the two HEXTE clusters were taken as free parameters, owing to
uncertainties (\ltsim 5\%) in the HEXTE deadtime measurement.  We then
verified that the fitted relative normalizations were consistent with
those derived from fits to the Crab. 

No single component model could adequately fit the full 2.5-250\,keV
spectrum.  In particular, excess emission above a power law at low
energies (\ltsim10\,keV), Fe-K line emission, and (for the more
significant observations) a cutoff at high energies (\gtsim35\,keV), are all required.
Several complex models fit the spectra with acceptable $\chi ^2$.  For
example, the broad-band continuum (above the Fe-K line) could be fit
by any of a broken power law, a power law with an exponential cutoff
above \aprx 30\,keV, or a Comptonized spectrum (\cite{Sun80}).
Further, a black body, a disk (multicolor) black body (see
e.g. Mitsuda \etal 1984 \nocite{Mak84}), and a thermal bremsstrahlung
model were all adequate to reproduce the low energy excess.

Because it provided the best fits and is simple in form, we
concentrated on a model made up of a power law with an exponential
cutoff at high energies ($E^{- \Gamma}$, $E \leq E_{cut}$;
$E^{- \Gamma}\times e^{-(E-E_{cut})/E_{fold}}$, $E > E_{cut}$) plus a
Gaussian Fe-K line and a disk black body.  Low energy absorption due
to intervening interstellar and/or local gas was also allowed.

Fits to the individual pointings showed no evidence for spectral
variability prior to the rapid dimming of the source which began
around MJD 50929.  Spectra after this time may have been slightly
softer, but the change was not highly significant. We therefore added
all the data prior to that date to obtain the most significant
spectrum. The PCA data below 25\,keV are of extremely high statistical
significance, so that uncertainties are dominated by small systematics
in the instrument response matrix.  We applied systematic errors to
the spectrum of between 0.5 and 2\% of the count rate per channel,
inferred from fits to the Crab.  In addition, we fixed the Gaussian
width of the Fe-K line at 0.1\,keV ($\sigma$), narrow compared to the
PCA energy resolution.

\begin{table}
\caption{\label{t_spec}Spectral fits to the \xray\ burster \saxj\ and three black
hole candidates.}
\begin{minipage}{\linewidth}
\renewcommand{\thefootnote}{\thempfootnote}
\begin{tabular}{lllll}
\hline\hline
 	  & \saxj & Cyg~X-1 & \onee & \Grs \\ \hline
$\rm N_H$ ($\rm \times 10^{22}\,cm^{-2}$)& $\rm 0.37^{+0.26}_{-0.21}$ & $\rm 0.3 \pm 0.2$ & 
$\rm 8.6\pm 0.6$ & $\rm 1.3\pm 0.3$ \\

$\rm T_{in}$\footnote{Temperature at the disk inner edge.} (keV) & $0.99^{+0.08}_{-0.02}$ & $\rm 1.4 \pm 0.1 $  & 
$\rm 1.15 \pm 0.30 $ & $\rm 1.4 \pm 0.2$ \\

$\rm K_{dbb}\footnote{Normalization of the disk blackbody
component: $K_{dbb}=(R_{in}/D_{10})cos\theta$. $R_{in}$ is the disk inner radius in
km, $D_{10} $ is the distance in units of 10\,kpc, and $\theta$ is the
angle of the disk.}$ & $13^{+2}_{-5}$ & $\rm 18 \pm 4$ & 
$\rm 1.2^{+0.8}_{-0.5}$& $\rm 0.87^{+.99}_{-.30}$ \\

$\rm E_{Fe}$ (keV) & $6.8^{+0.1}_{-0.2}$ & $\rm 6.47^{+0.06}_{-0.09}$ &
$\rm 5.7 \pm 0.7$\footnote{Additional systematic uncertainties apply due to the
subtraction of diffuse Galactic plane Fe-K emission.}  & $\rm 6.28^{+0.12}_{-0.07}$\footnotemark[\value{mpfootnote}]\\

$\rm EW_{Fe}$\footnote{Fe-K line equivalent width.}\addtocounter{mpfootnote}{-1} (eV) & $52^{+9}_{-8}$ & $\rm 90^{+13}_{-8}$ & 
$\rm 19^{+19}_{-14}$\footnotemark[\value{mpfootnote}] & $77 \pm 9 $\footnotemark[\value{mpfootnote}]\addtocounter{mpfootnote}{1}\\

$\Gamma$\footnote{Power law photon index.} & $\rm 1.86 \pm 0.01$ & $\rm 1.488 \pm 0.014 $ & 
$\rm 1.53 \pm 0.06$ & $\rm 1.54 \pm 0.04$\\

$\rm E_{cut}$ (keV) & $\rm 34 \pm 4$ & $\rm 34 \pm 5 $ &
$19.1 \pm 3.5$ & $\rm 30^{+20}_{-10}$\\

$\rm E_{fold}$ (keV)& $127^{+22}_{-13}$  & $\rm 230 \pm 20$ & 
$\rm 116 \pm 20 $& $\rm 185^{+75}_{-50}$ \\

$\rm L_{1-20\,keV}$\footnote{Luminosity ($10^{36}$\,\lumin).  Assuming distances
of \saxj: 4\,kpc, Cyg~X-1: 1.9~kpc, \onee: 8.5\,kpc, and \Grs:
8.5\,kpc.} & 2.5 & 5.2 & 7.0 & 5.7\\

$\rm L_{20-200\,keV}$\footnotemark[\value{mpfootnote}] & 1.9 & 11 & 15 &  12\\
\hline
\end{tabular}
\end{minipage}
\end{table}

\section{Results}

\begin{figure}
\caption{The spectrum of \saxj\ between MJD 50914 and
50928. The upper panel shows the measured and best fit counts spectra
(data points and histograms) and the inferred incident photon spectrum
(smooth curve).\label{f_spec}}  
\centerline{\epsfig{file=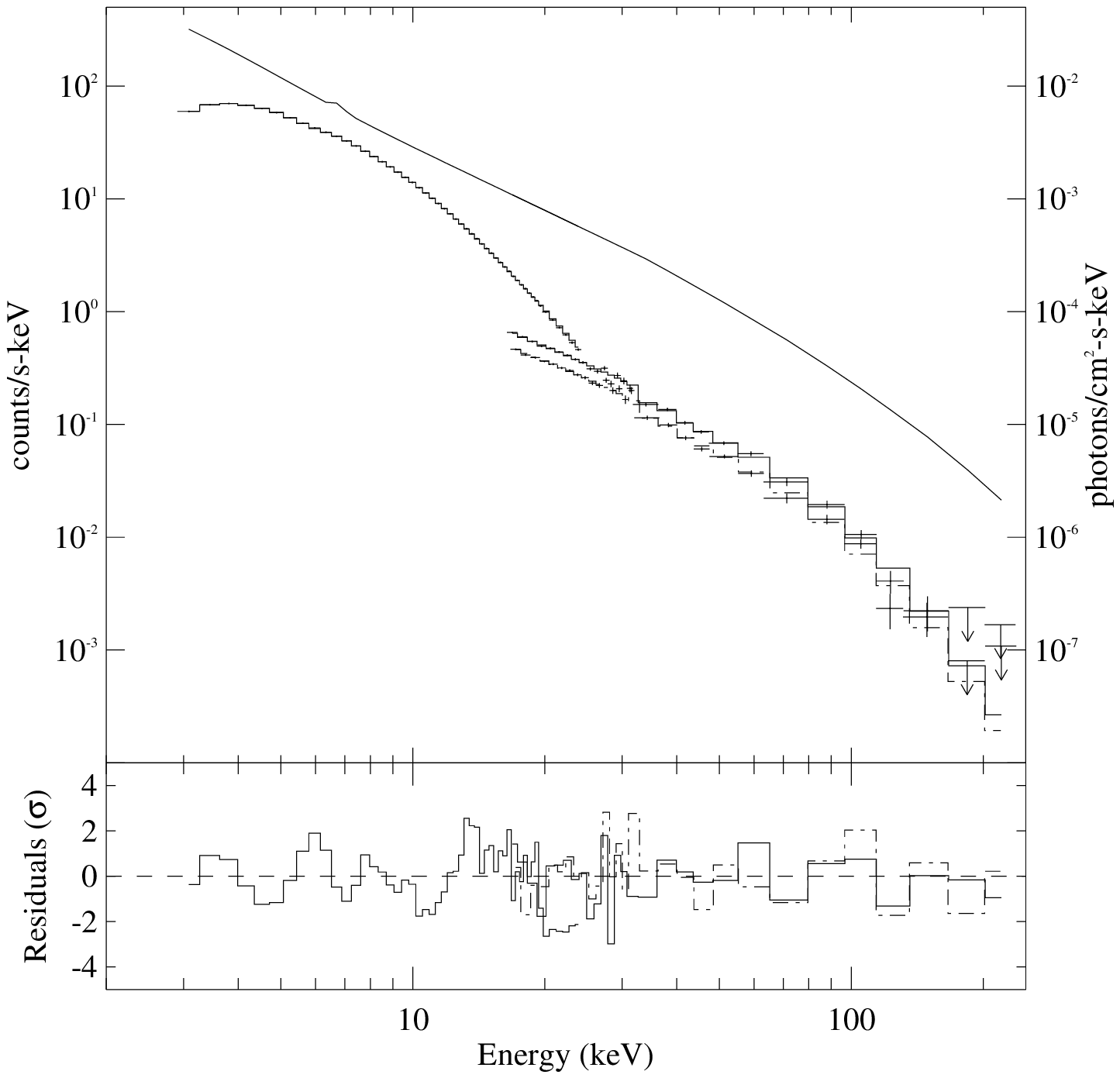,width=4.5in}}
\end{figure}

\begin{figure}
\caption{The ratio of the PCA counts spectra of \saxj\ and
the Crab Nebula and Pulsar.  The data have been arbitrarily normalized
to 1 at 10\,keV.\label{f_crabrat}}
\centerline{\epsfig{file=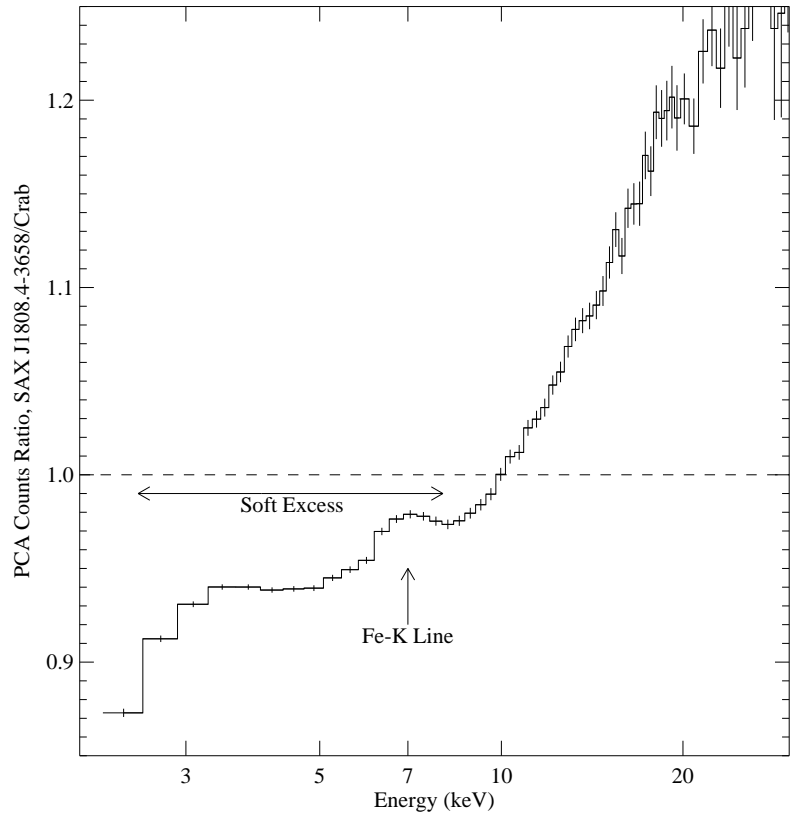,width=4.5in}}
\end{figure}

The total spectrum for observations between MJD 50914 and 50928
together with the best fit model and the inferred incident spectrum of
the form described above are shown in Figure~\ref{f_spec}. The model
parameters are listed in Table~\ref{t_spec}. Although we included
systematic uncertainties in the PCA data, we were concerned that the
residuals to a simple power law model may have been caused by small
errors in the response matrix.  So, to verify our results, we divided
the PCA data by a Crab Nebula and Pulsar spectrum (provided by
K. Jahoda of the PCA team) which was obtained on MJD 50914.  The ratio
is shown in Figure~\ref{f_crabrat}.  Because the Crab is a very bright
source, the statistics of its spectrum are negligible in the ratio.
Since the spectrum of \saxj\ is similar in overall shape, the ratio
gives a fairly matrix-independent picture of features which differ
from the Crab.  In particular, the overall positive slope is due to
the harder power law in \saxj, and the soft excess and an iron line
are clear below 10\,keV.  While we remain cautious of the physical
interpretation of the soft excess as a disk black body and the
quantitative iron line model parameters, we are confident that both
the soft excess and the iron line are present.

Figure~\ref{f_crabrat} also shows that the resemblance to the Crab is
at best superficial.  In contrast to the power law fits to the
individual observations made by Gilfanov \etal\ (1998)\nocite{Glf98},
we find strong deviations from such a simple model ($\chi^2_{red} =
22.4$ with 94 degrees of freedom (DOF)), even in the limited range of
3--100\,keV which they used.  The most important excursions are at low
and high energies where we find a soft excess and an exponential
cutoff respectively.  The fit in Table~\ref{t_spec} for \saxj\ had
$\rm \chi^{2}_{red} = 1.65$ for 99 DOF (2.5--250\,keV).  If we
disallow the exponential cutoff and refit all the other parameters, we
find $\rm \chi^{2}_{red} = 4.34$ for 101 DOF.  Applying an $F$-test
(\cite{Bev69}) to the ratio of the improvement in $\chi^{2}$ and
$\chi^{2}$ of the full model (divided by 2 and 99 DOF, respectively),
we find that the probability that the cutoff is unnecessary is only $7
\times 10^{-22}$.

We also analyzed the PCA spectra in eight bins by orbital phase. Due
to the limited amount of data, each phase bin samples the different
parts of the decaying lightcurve to different degrees.  To compensate
for this effect, we fit an exponential decay to the lightcurve and
looked for additional variations.  We find an apparent intensity
variation with orbit, in agreement with Chakrabarty \& Morgan (1998b):
roughly sinusoidal, with an amplitude of 2\%, and a minimum when the
neutron star is behind its companion. The modulation is several times
larger than could be explained by uncertainties in the background
model. We find no differences in the spectrum with phase: the disk
blackbody and power law components share the extinction equally, and
the absorption column is unchanged.  This may imply that the
modulation is due to Compton scattering in a thin, ionized intervening
medium (e.g. the ablated wind suggested by Chakrabarty \& Morgan
(1998b)), which would scatter photons out of the line of sight nearly
independently of energy.  Alternately, the modulation could be an
artifact of the limited amount of data, caused by random correlations
between the orbital phase and small variations in the accretion rate:
the total amount of data available is only about 14 orbits.
 
\section{Discussion}

\subsection{Comparison to Type I Burst Sources}

Prior to this work, the only bursters with nearly simultaneous
coverage of the \xray\ and hard \xray\ bands during episodes of hard
outburst were Cen~X-4 and 4U~1608-52.  In 1979, Cen~X-4 was observed
with {\em Hakucho} (1.5--30\,keV) (\cite{Mat80}), {\em Ariel 5}
(3--6\,keV) (\cite{Kal80}), and {\em Prognoz 7} (13-163~keV)
(\cite{Bou84}).  Bouchacourt \etal\ (1984) fit the {\em Prognoz 7}
spectrum to a hard power law (photon index fixed: $\Gamma \equiv 1$)
times an exponential cutoff with a folding energy of \aprx 50\,keV --
not unlike the \rxte\ \saxj\ spectrum.  They also noted that the low
energy extrapolation of the {\em Prognoz 7} data fell below the
simultaneous {\em Ariel 5} spectrum, suggesting soft excess emission,
although this could be due to the true power law differing from $\rm
E^{-1}$.  Zhang \etal\ (1996a) \nocite{Zha96a} used the \ginga\
spectrum (\cite{Yos93}) from the middle of a \aprx 170 day (1991 June
-- December) outburst to constrain the low energy portion of the
4U~1608-52 spectrum observed by BATSE.  The \ginga\ data showed a
power law with photon index of $\rm -1.75 \pm 0.01 $, and together
with the BATSE data the spectrum was fit by a Comptonized model
(\cite{Sun80}) with $\rm kT = 23$\,keV and optical depth 4.4 (for a
spherical geometry).  For \saxj, we find nearly identical values of
$\rm 22.0^{+1.6}_{-0.8}\,keV$ and $\rm 4.02^{+0.11}_{-0.15}$ for the
electron temperature and optical depth.  Thus, the broad band spectra
of all three well-studied hard state burst sources are quite similar.

Hard \xray\ emission from other bursters has been observed (without
concurrent soft \xray\ coverage) with BATSE and SIGMA (see Barret,
McClintock, \& Grindlay 1996 and references therein).  When these data
are fit to broken power law spectra, photon indices of between 2.5 and
3 above the break are generally found.  Fitting the HEXTE data above
40\,keV gives an index of $\rm 2.4 \pm 0.1$ for \saxj, in reasonable
agreement with the high energy spectra of other low state bursters.

\subsection{Comparison to Black Hole Candidates}

Historically, hard emission extending to \gtsim 100\,keV from \xray\
binaries has been considered a ``black hole signature'',
distinguishing black hole sources from low magnetic field neutron
stars.  Clearly with several burst sources emitting hard \xrays, the
situation is not so simple.  However, it may still be possible to
separate these neutron stars by other properties of their
emission. For example, Barret, McClintock, \& Grindlay (1996)
\nocite{Bar96} noted that neutron stars only produce hard tails when
the tails dominate the emission, as opposed to black holes, which can
produce hard tails along with very bright ultrasoft emission.  They
also noted that the luminosity when the hard emission dominates can be
higher for black holes than for neutron stars.  \saxj\ is consistent
with a neutron star by both these criteria, showing only a very weak
soft component and having a modest luminosity ($\sim 5\times 10^{36}$
ergs s$^{-1}$ assuming a distance of 4\,kpc).  Given that distances
are often poorly known and that black hole candidates can also be seen
at low luminosities, luminosity alone is an insufficient means to
distinguish between source types.

To see if the spectrum alone can serve the purpose, we fit \rxte\
observations of the persistent low state black hole candidates (BHCs)
Cyg~X-1 (1998 April), \onee (1996 March), and \Grs\ (1996 August) to
the same model used for \saxj.  These fits are listed in
Table~\ref{t_spec}.  One feature of the BHC spectra clearly differs
from the bursters: the photon indices are significantly harder in the
BHCs.  The BHC's power laws are between 1.4 -- 1.6, while the \saxj\
index is 1.86. From \ginga\ observations, 4U~1608-522 had an index of
1.75 during the period of hard emission described above.  Thus, from
this limited sample, the broad-band spectrum separates the BHCs and
neutron stars via the hardness of the power law.  Given the realities
of intercalibrating multiple instruments and the modest separation of
the indices, this characteristic may be difficult to use for
comparisons between different observatories.  However, it will clearly
be useful for observations made with a single instrument complement or
between instruments which are well intercalibrated.

\section{Conclusions}

The spectrum of the unique 401~Hz pulsar and type~I burster \saxj\
during its recent outburst was quite hard. The photon spectral index
was 1.86, with a slow cutoff at high energies.  There was clear
evidence of excess soft emission and a weak Fe-K line.  These \rxte\
observations have provided the highest quality broad-band spectrum of
a suspected \xray\ burster during a period of hard emission.  The
spectrum is in good agreement with previous observations of other
bursters made over more limited energy bands and with lower
statistical significance. By comparing to observations of low state
BHCs, we find that BHCs and bursters can be distinguished by the slope
of their power law emission.  In particular, the photon indices of
bursters are greater (i.e. softer) by about 0.3, even though the
overall spectral shapes are quite similar.

\acknowledgements

Thanks to Keith Jahoda for his help with the PCA response matrix and
for providing the Crab data.  We would also like to acknowledge the \rxte\
Science Operations Center for scheduling these Target of Opportunity
observations. This work was supported by NASA grant NAS5-30720.


\newpage

\end{document}